\documentclass[conference]{IEEEtran}
\IEEEoverridecommandlockouts
\usepackage{cite}
\usepackage{amsmath,amssymb,amsfonts}
\usepackage{graphicx}
\usepackage{textcomp}
\usepackage{xcolor}
\def\BibTeX{{\rm B\kern-.05em{\sc i\kern-.025em b}\kern-.08em
    T\kern-.1667em\lower.7ex\hbox{E}\kern-.125emX}}
\usepackage{cite}
\usepackage{amsmath,amssymb,amsfonts}
\usepackage{graphicx}
\usepackage{textcomp}
\usepackage{xcolor}
\usepackage{units}

\usepackage{pgfplots}
\usepackage{tikz}
\usepackage{algorithm}
\usepackage{algpseudocode} 

\def\BibTeX{{\rm B\kern-.05em{\sc i\kern-.025em b}\kern-.08em
    T\kern-.1667em\lower.7ex\hbox{E}\kern-.125emX}}
\usetikzlibrary{patterns}
\begin{document}


\newcommand{\note}[1]{\textcolor{red}{[#1]}}

\algnewcommand{\Inputs}[1]{%
\State \textbf{Inputs:}
\hspace*{\algorithmicindent}\parbox[t]{.8\linewidth}{\raggedright #1}
}

\algnewcommand{\Initialize}[1]{%
\State \textbf{Initialization:}
\Statex \hspace*{\algorithmicindent}\parbox[t]{.8\linewidth}{\raggedright #1}
}

\algnewcommand{\Output}[1]{%
  \State \textbf{Output:}
  \hspace*{\algorithmicindent}\parbox[t]{.8\linewidth}{\raggedright #1}
}

\algnewcommand{\Update}[1]{%
  \State \textbf{Update:}
  \hspace*{\algorithmicindent}\parbox[t]{.8\linewidth}{\raggedright #1}
}

\title{Resource Allocation with Stability Constraints of an Edge-cloud controlled AGV}


\author{
\IEEEauthorblockN{Shreya Tayade\IEEEauthorrefmark{1},
Peter Rost\IEEEauthorrefmark{2},
Andreas Maeder\IEEEauthorrefmark{3} and 
Hans D. Schotten\IEEEauthorrefmark{4}}  
\IEEEauthorblockA{\IEEEauthorrefmark{1}Intelligent Networks Research Group, German Research Center for Artificial Intelligence, Kaiserslautern, Germany\\
Email: \{Shreya.Tayade, Hans\_Dieter.Schotten\} @dfki.de}
\IEEEauthorblockA{\IEEEauthorrefmark{2}Communications Engineering Lab, Karlsruhe Institute of Technology, Germany \\ Email: peter.rost@kit.edu}
\IEEEauthorblockA{\IEEEauthorrefmark{3}Nokia Bell Labs,
Munich, Germany\\
Email: andreas.maeder@nokia-bell-labs.com}
 
\IEEEauthorblockA{\IEEEauthorrefmark{4} Wireless Communication and Navigation, Technical University of Kaiserslautern,  Germany \\ Email: schotten@eit.uni-kl.de}
}

\maketitle

\begin{abstract}
The paper proposes Resource Allocation (RA) schemes for a closed loop feedback control system by analysing the control-communication dependencies. We consider an Automated Guided Vehicle (AGV) that communicates with a controller located in an edge-cloud over a wireless fading channel. The control commands are transmitted to an AGV and the position state is feedback to the controller at every time-instant. A control stability based scheduling metric 'Probability of Instability' is evaluated for the resource allocation. The performance of stability based RA scheme is compared with the maximum SNR based RA scheme and control error first approach in an overloaded and non-overloaded scenario. The RA scheme with the stability constraints significantly reduces the resource utilization and is able to schedule more number of AGVs while maintaining its stability. Moreover, the proposed RA scheme is independent of control state and depends upon consecutive packet errors, the control parameters like sampling time and AGV velocity. Furthermore, we also analyse the impact of RA schemes on the AGV's stability and error performance, and evaluated the number of unstable AGVs. 
\end{abstract}


\section{Introduction}
Introduction of tactile internet have initiated use-cases like tele-surgery and tele-medicine in E-health sector, Machine to Machine (M2M), Human to Machine (H2M) communication in the automation industry and self driving cars in automotive industry. The tactile internet applications are driven by closed loop feedback controls that have stringent latency requirements constituting the Round Trip Time (RTT) to \unit[1]{ms} \cite{electronics10172171}. The promising solution to support growing demands of reliability and ultra low latency for a real-time feedback control is an Edge cloud. Edge cloud enables computation of complex processing at the edge unlike the centralized cloud, hence reducing the latency \cite{8403496}.

\begin{figure*}[t!]
\centering
\includegraphics[width=0.8\textwidth]{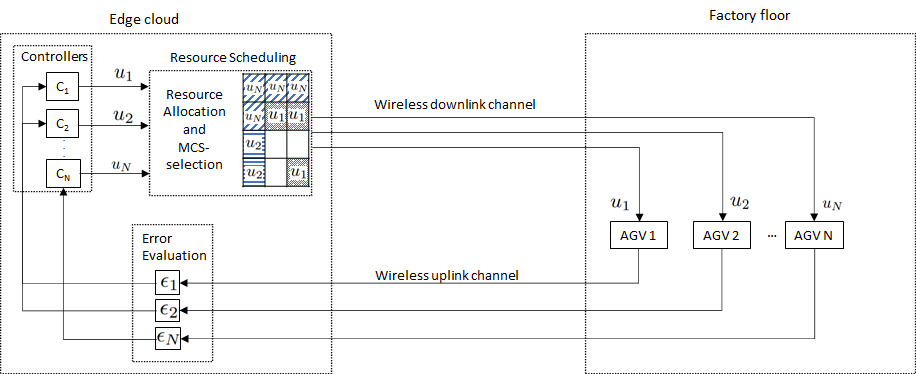}
\caption{System Model}
\label{fig:ch6:1}
\end{figure*}


Moreover, unlike conventional cellular communication, where the throughput, data rate are the Key Performance Indicators (KPI), the performance of control application is determined with the stability and control state error. In an edge cloud based Wireless Network Control System (WNCS) the controller sends the control commands via downlink wireless channel and receives the state feedback via uplink. The stability and control state error depends up on the freshness of control commands also known as Age of Information (AoI) \cite{AOI, binhan}. A fading channel may cause consecutive control packet loss hence increasing the AoI as the control information will be outdated. Therefore, the channel quality is crucial to maintain the stability of an AGV. 

Additionally, in an industrial control short packets are transmitted between the controllers and actuators either periodically or are event-triggered. The error probability to transmit short packets is higher due to finite-block length. High error probability may cause instability due to loss of control packets that leads to higher AoI. The stability of control system is significantly dependent on the channel condition, and control packet transmission strategies. The control packet transmitted with lower Modulation and Coding Scheme (MCS), with more radio resources will reduce the error probability and can retain stability. However, allocating more resources than required to maintain the stability may cause resource wastage. Therefore, the control aware RA techniques should be adopted to efficiently utilize resources. The traditional RA strategies like proportional fair, Round Robin and maximum SNR focuses on maximizing the data rates by considering the wireless channel quality, and providing fairness. However, for industrial control the stability and control state error is the priority as to the system throughput. Therefore, new control aware scheduling metrics must be evaluated to maintain control system stability and efficiently use the resources. 

\footnote{This is a preprint, the full paper has been published in $27^{th}$ VDE ITG European Wireless Conference, \copyright 2022 VDE. Personal use of this material is permitted. However, permission to use this material for any other purposes must be obtained from the ITG VDE society at itg@vde.com}

\subsection{Related work} Channel aware encoding-decoding strategies to mitigate the effect of noisy and lossy communication link are proposed in \cite{1310480NoisyChannel}. The impact of packet error rate, delay, channel noise and lossy communication links on the performance of control system is studied in \cite{1310480NoisyChannel,1310461comm_constraints,gracia,inbook,5409530,898792,1184679,998034}. The research in \cite{fbl} focuses on evaluation of control aware optimal coding rate for transmission that increases resource efficiency. Few control aware RA schemes are also proposed in \cite{898792,5409530,8681425, 8011303, mikhail, Scheuvens, Gatsis}. A promising control state aware RA for multi-connectivity is also presented in \cite{Scheuvens}  to increases the availability.  The authors in \cite{mikhail} present the comparison of  control aware scheduling metrics for RA of an Linear Time Invariant (LTI) control system in the Uplink. A network induced error and the predicted control error first approach is compared with the traditional control error first approach \cite{898792}. The predicted control error approach, predicts the control error with control system parameters and is independent of time and network errors. Therefore, the scheduler is only applicable for LTI system, however, in practice the control loops are often complex, time-varying, non-linear and in-homogeneous in nature. 

\subsection{Contribution and outline of the paper}
In this paper we propose the control aware RA schemes for a non-linear pragmatic edge cloud based AGV control system and investigate its stability performance. The probability of instability is evaluated considering the channel conditions and control system stability. We compare control aware scheduling metrics with the conventional SNR as scheduling metric. 
A control and communication system model is presented in section~\ref{sec:ch6:1}. The MCS selection and resource allocation to transmit the control packet is described in section~\ref{subsec:ch6:15}. The control aware scheduling metric and the proposed resource allocation algorithm are presented in \ref{sec:ch6:10} and \ref{sec:ch6:15} respectively. The results and conclusions are discussed in section~\ref{sec:ch6:20} and \ref{sec:ch6:25} respectively.

\section{System Model} \label{sec:ch6:1}
\subsection{AGV control system} \label{subsec:ch6:5}
The system consists of AGVs controlled by an edge-cloud based independent controllers $C_i$, where $i = 1\cdots N$ represents the $i^{th}$ AGV as shown in  Fig.~\ref{fig:ch6:1}. The controller $C_i$ generate control command $u_i(k)$ at $k^{th}$ time-step at an interval of control system sampling time $T_s$, and is sent to an AGV in the downlink. The control command consists of an instantaneous velocity and acceleration for driving an AGV on the desired reference track. The desired reference track $X_{r,i}$= $\left(x_{r,i}(k), y_{r,i}(k), \theta_{r,i}(k) \right)$  is determined at every time step $k$, where $(x_r,y_r)$ represents the position coordinates and $\theta_r$ is the orientation of an AGV. The actual track that AGV follows is represented by $X_{c,i}(k)$ = $\left(x_{c,i}(k), y_{c,i}(k), \theta_{c,i}(k) \right)$. The subscript $r$ represent reference position and $c$ represents current AGV position of $i^{th}$ AGV at $k^{th}$ time-instant. An AGV attains new position $X_{c,i}(k+1)$ on applying the control commands as
\begin{equation}\label{eq:ch6:10}
    X_{c,i}(k+1) =  X_{c,i}(k) +  T_s \cdot J(k) \cdot  u_i(k)
\end{equation}
where $J(k)$ is transformation matrix. 
The attained position of an AGV $ X_{c,i}(k+1)$, is feedback to the controller via uplink channel. The error $\epsilon_{i,k}$ is evaluated as the deviation of an AGV from the reference track as shown in Fig.~\ref{fig:ch6:1}. The controller generates new control commands $u_i(k+1)$ corresponding to the error $\epsilon_{i,k}$. For the sake of brevity an uplink channel is assume to be perfect. The details about controller is described in \cite{AGVSCC, tayade2021error, Kanayama1990}.

\subsection{Communication system}
AGVs are uniformly distributed in a circular area of radius $r$ from an edge-cloud. The control command $u_i$ is represented by $D_i$ data bits at every scheduling time instant $T_s$. The sampling time of control system is assumed to be same as the scheduling time instant. A Rayleigh fading channel with path-loss exponent $\beta$ is considered for downlink communication. At every time instant the distance $d_{i,k}$ of an AGV from the edge cloud is evaluated. The received SNR $\gamma_{i,k}$ is known considering the path-loss and time correlated channel as
\begin{align}
    \gamma_{i,k} = P_{tx} \times \left( \frac{\lambda}{4\pi d_0}\right)^2 \times \left( \frac{d_0}{d_{i,k}}\right)^\beta \frac{\lvert h_{i,k} \rvert^2}{N_o BW}
\end{align}
where $P_{tx}$ is the transmit power, $d_0$ is the reference distance, $d_{i,k}$ is the distance between the $i^{th}$ AGV and an edge cloud at $k^{th}$ time-step, $h_{i,k}$ is the channel gain, $\lambda$ is the wavelength corresponding to carrier frequency and $N_o$ is the noise. The AGVs are scheduled on the available bandwidth $BW$ for downlink communication that corresponds to $N_{RB}$, the total number of Resource Blocks (RBs) per $T_s$ and $N_{rb,i}$ is the number of RBs allocated to an $i^{th}$ AGV at every scheduling time instant. The Modulation and Coding Scheme (MCS) is selected at every $T_s$ to encode the $D_i$ data bits with Coding rate $R_i$ and modulation scheme $M_i$. The error probability $P_e^{i,k}$ is
\begin{align}
    P_e^{i,k} = f(\gamma_{i,k}, R_i, M_i),
\end{align}
and depends on the $\gamma_{i,k}$ and the selected MCS scheme. 
The expected error probability for $i^{th}$ AGV will vary with the modulation and coding scheme. For constant MCS, the expected error probability is 

\begin{align}
 \mathbf{E}_\gamma \lbrace P_e^{i} (R_i,M_i)\rbrace = \int_0^\infty P_e^{i}(\gamma_{i}, R_i,M_i) f_{\gamma_i} (\gamma_i) d\gamma_i, 
\end{align} where $f_{\gamma_i}(\gamma_i)$ is the PDF for SNR of i-th AGV given as
\begin{align}
    \gamma_i &= \lvert h_i \rvert ^2 \gamma_{b,i} \\
    f_{\gamma_i} &= \frac{1}{\gamma_{b,i}} e^{\left(\frac{-\gamma_i}{\gamma_{b,i}}\right)} \\ 
    \gamma_{b,i} &= \frac{P_{tx} \times \left( \frac{\lambda}{4\pi d_0}\right)^2 \times \left( \frac{d_0}{d_{i,k}}\right)^\beta}{N_o BW}
\end{align}

The success or failure of current transmission is identified by $\delta_{i,k}$. The $\delta_{i,k}$ determines the number of consecutive packet losses in the past, for $i^{th}$ AGV at $k^{th}$ timestep. 

\begin{align}\label{eq:RA:50}
    \delta_{i,k} &= 0 \quad \text{Successful Transmission}  \nonumber \\
    \delta_{i,k} &= \delta_{i,k-1} + 1 \quad \text{Unsuccessful Transmission}
\end{align}

$\delta$ represents the AoI of control packets in the downlink as it determines the number of consecutive control commands not successfully received by an AGV.  

\subsection{Resource allocation and MCS selection} \label{subsec:ch6:15}
The MCS selection depends upon the allocated RBs for transmission in a resource limited regime. We assume a LTE resource grid with each RB consisting of 12 sub-carriers and 7 OFDM symbols. If $N_{rb,i}$ is the amount of RBs allocated to $i^{th}$ AGV and $M_i$ is the modulation scheme, the permissible coding rate is 
\begin{align}
     R_i =   \frac{D_i}{12 \times 7 \times N_{rb,i} \times \log_2(M_i)}.
\end{align}

Therefore, the control packet of size $D_i$ bits can be encoded with coding rate $R_i$ for modulation scheme $M_i$, if the allocated RBs $N_{rb,i}$ is  
\begin{align} \label{eq:ch6:100}
    N_{rb,i}  \geq  \lceil \frac{D_i}{12 \times 7 \times \log_2(M_i) \times R_i} \rceil.
\end{align}

\pagebreak 
\section{Metrics for Resource Allocation} \label{sec:ch6:10}
The resource allocation schemes for cellular networks focuses on increasing the system throughput and fairness, hence, more resources are allocated to the user with maximum SNR, simultaneously ensuring fairness. In industrial control the stability and the control error are crucial factors as compared to the throughput, considering the short data packet size. In order to maintain the stability of AGVs, we present the control dependent scheduling metrics for allocating resources. 


\subsection{Maximum control error}
In this RA scheme, the RBs are allocated considering the control error. The control error $\epsilon_{i,k}$ in AGV system is the deviation of $i^{th}$ AGV from the reference track $X_{r,i}(k)$ at time step $k$. The maximum deviation of an AGV allowed from its desired reference path is the error threshold $\epsilon_{th}$. The error at every time-step should be less than the error threshold i.e. $\epsilon_{i,k} < \epsilon_{th}$. The error is evaluated as 
\begin{align}
    \epsilon_{i,k} = \sqrt{(X_{r,i}(k) -X_{c,i}(k))^2 + (Y_{r,i} (k) -Y_{c,i}(k))^2}.
\end{align} 
In this resource allocation scheme, the AGV with maximum control error is given the priority and more resources are allotted to successfully transmit the control packet in the current scheduling time instant. 

The priority user is 
\begin{align}
    \textbf{Priority User} = \min_i \: \lvert \epsilon_{th} -  \epsilon_{i,k} \rvert,
\end{align}
and the control error is independent of channel conditions. It is evaluated from the position feedback received by the edge cloud controller in the uplink.


\subsection{Maximum SNR} In this RA scheme the objective is to maximize the overall system throughput. The AGV with maximum SNR is given priority for successful downlink transmission and higher data rates. The RA is independent of the control system parameters like control error and stability of an AGV. More resources are allocated to an AGV with higher SNR as

\begin{align}
     \textbf{Priority User} = \max_i \: \gamma_{i,k}.
\end{align}

\subsection{Probability of instability}
The AGV is unstable if $\delta_{i,k} >  n_{max}$, where $n_{max}$ is the maximum number of consecutive packet loss that an AGV control system can sustain without being unstable \cite{AGVSCC}. The stability of an AGV depends upon $\delta_{i,k}$, higher the number of consecutive packet error, higher is the probability of an AGV becoming unstable.  If an AGV transmits with constant MCS scheme, the probability of $n_{max}$ consecutive packet error $P_{us}^i$ for $i^{th}$ AGV can be evaluated as 

\begin{align}
   P_{us}^{i}(1,\cdots, n_{max}) = P_{bb}^{n_{max}-1} \times \mathbf{E}_\gamma \lbrace P_e^{i} (R_i,M_i)\rbrace
\end{align}
where $P_{bb}$ is (back-back error probability),
\begin{align} \label{eq:ch6:15}
    P_\text{bb} = \frac{1-\mathbb{E}_\gamma \lbrace P_e^i(R_i,M_i)\rbrace}{\mathbb{E}_\gamma \lbrace P_e^i(R_i,M_i)\rbrace} \left[ Q(\theta,\rho\theta) -Q(\rho\theta,\theta)] \right], 
\end{align}
where 
\begin{align}
\theta &= \sqrt{ \frac{-2\log(1-\mathbb{E}_\gamma \lbrace P_e^i(R_i,M_i)\rbrace)}{1- \rho^2}}, \\
\rho &= \mathbf{J}_0(2\pi f_d T_s),
\end{align}
and $P_e^i$ is the average block error probability that depends upon MCS scheme. 

The value of $\delta_{i,k}$, determines the number of future opportunities still available to successfully transmit the control packet to prevent instability. The control packet must be successfully received within next ($n_{max}-\delta_{i,k}$)  transmissions. If $\delta_{i,k} = 0$, $n_{max}$ future opportunities are available to successfully transmit the packet. The higher the value of $\delta_{i,k}$, few transmission opportunities are available. The AGV is unstable if the next ($n_{max}-\delta_{i,k}$) transmission fails. 

The probability of instability is therefore the
probability of $n_{max}-\delta_{i,k}$ consecutive packet error 
\begin{align}
P_{us}^{i,k} &= P_{us}^{i}(1,\cdots, n_{max}-\delta_{i,k}) \\
P_{us}^{i,k} &=  P_{bb}^{n_{max}-\delta_{i,k}- 1} \times \mathbf{E}_\gamma \lbrace P_e^{i} (R_i,M_i)\rbrace, 
\label{eq:ch6:200}
\end{align}
where $\delta_{i,k}$ is the measure of consecutive block error up to $k-1$ time-step evaluated in \eqref{eq:RA:50}. 
In this resource allocation scheme, the AGV that has higher consecutive block error i.e. higher probability of instability is given the priority as
\begin{align}
     \textbf{Priority User} = \max_i \: P_{us}^{i,k}.
\end{align}


\section{Proposed Resource Allocation } \label{sec:ch6:15}
\subsection{Error probability threshold and instability}
The stability of an AGV depends upon the number of consecutive packet errors \cite{AGVSCC}. If the control packets are transmitted with higher frequency in the downlink, i.e. with lower sampling time, the channel correlation is higher over two consecutive transmission. If the channel correlation is higher, more is the probability of back to back failure, hence higher is the probability of instability. 

Furthermore, the expected block error probability is higher if higher MCS scheme is selected. The probability of instability is the probability of $n_{max}$ consecutive packet failure, given $\delta_{i,k} = 0$. Therefore, if block error probability is high the probability of instability will also increase as shown in \eqref{eq:ch6:200}. The Block Error Rate (BLER) for different MCS scheme over SNR, for LTE is available in a lookup table. The MCS scheme is selected such that the error probability is below the threshold error probability for every transmission as
\begin{equation}\label{eq:ch6:201}
MCS_i(M_i, R_i) =  \lbrace MCS(M,R)| P_e^i(\gamma_{i,k}, R_i,M_i) < P_e^{th}\rbrace, 
\end{equation} 
$\forall i = 1\cdots N$, where $P_e^{th}$ is the threshold error probability.
 
The reliability requirement for closed loop control may reach up to the order of $10^{-9}$, and hence requires extremely low threshold error probability for successful transmission. However, some control loops can maintain the stability even with lower reliability, therefore stringent constraints on $P_e^{th}$ may result in to inefficient resource utilization. Moreover, as the reliability requirements vary for different control systems, we impose constraints on its stability, i.\,e. the $P_{us}^{i,k} \leq 10^{-9}$. Therefore, we evaluate the inter-dependencies between the $P_e^{th}$ and stability of the system. The probability of instability for $i^{th}$ AGV is evaluated from \eqref{eq:ch6:200}. 
As the $\delta_{i,k}$ increases, less number of attempts are left to successfully transmit the control packet, hence $P_{us}^{i,k}$ is high. Moreover if $P_e^{i}$ is high, i.e. higher MCS scheme selected, then higher is the $P_{us}^{i,k}$. Therefore, we evaluate the $P_e^{th}$, for every AGV at every scheduling time instant so that the optimal MCS scheme is selected such that $P_{us}^{i,k}(n_{max} - \delta_{i,k}) \leq 10^{-9}$.

The $P_e^{th}$ is 
\begin{align} \label{eq:ch6:500}
P_e^{th} = \lbrace \mathbf{E}_\gamma \lbrace P_e^{i} (R_i,M_i)\rbrace | P_{us}^{i,k}(n_{max} - \delta_{i,k}) \leq 10^{-9} \rbrace, \\
MCS_i(M_i, R_i) =  \lbrace MCS(M,R)| P_e^i(\gamma_{i,k}, R_i,M_i) < P_e^{th} \rbrace. \label{eq:ch6:501}
\end{align}
The $P_e^{th}$ is numerically evaluated for varying value of $\delta_{i,k}$. $P_e^{th}$ decrease if $\delta_{i,k}$ increases, as the AGV requires higher reliability for packet transmission. The AGV can tolerate few more packet loss if $\delta_{i,k} = 0$ hence the packet can be transmitted with higher MCS scheme, i.e. less resource consumption. 



\begin{algorithm}[h!] \label{alg:ch8:1} 
\caption{$P_{us}$ based resource allocation}
\begin{algorithmic}
\Initialize{ k=1, $\delta_{i,k}$ = 0, $N_{rb,i} = 0$; $\forall i = 1 \cdots N $, $P_e^{th} = 10^{-3}$ }\\
Evaluate: $P_e^{i,k}$ = $f(\gamma_{i,k}, MCS_{id})$  \Comment{Equal allocation}  \\ 
\For{($k = 1 \to k_{max}$)}
\State Evaluate:  $P_{us}^i(n_{max}-\delta_{i,k}) \forall i= 1 \dots N$\\
\For{$i = 1 \to N$} \\ 
\If {$\sum_i^N N_{rb,i} < N_{RB}$} \Comment{RBs available} \\
\State{Priority AGV:} 
\State{ $i_p$ = $ \text{arg}\underset{i= 1 \ldots N}{\max} (P_{us}^i(n_{max}-\delta_{i,k})$)}  \\ 

\State {Evaluate: $P_e^{th}$ as in \eqref{eq:ch6:500}} 
\State{Evaluate: MCS scheme}
\State{$(M_{i_p},R_{i_p})$ according to \eqref{eq:ch6:501}} \\
\State{Evaluate: $N_{rb,i_p}$ ~\eqref{eq:ch6:100}} \Comment{RBs required for ($M_{i_p},R_{i_p}$)} \\ 
\State{Assign $N_{rb,i_p}$ RBs to priority AGV}
\Update{$P_e^{i_p,k}$ = f($\gamma_{i_p,k}$,$M_{i_p},R_{i_p}$)}
\ElsIf{$\sum_i^N N_{rb,i} = N_{RB}$}
\State $N_{rb_i}$ = 0 
\Comment{No RBs available}  
\Update{$P_{e}^{i,k}$ = 1} 
\EndIf
\EndFor 
\Update{$\delta_{i,k}$, $P_e^{i,k}$,$N_{rb,i}  \forall i = 1\ldots N $} 
\State{AGV control and state evolution} 
\State{$X_{c,i}(k+1) =  X_{c,i}(k) +  T_s \cdot J(k) \cdot  u_i(k - \delta_{i,k} )$}
\Output{$\epsilon_{i,k}$} \Comment{Control error}
\EndFor
\end{algorithmic}
\end{algorithm}

\subsection{Algorithm}
The AGVs arrive with Poisson process and are active for certain time period to trace the desired reference track $X_{r,i}$. The RA algorithm assigns RBs at every scheduling time instant $k$, for all $N$ active AGVs. The error probability is evaluated assuming the RBs are equally distributed with $P_e^{th} = 10^{-3}$. The scheduling metric is evaluated for all the active AGVs and a priority user is chosen. The proposed $P_{us}^{i,k}$ based approach evaluates the probability of instability based on the consecutive packet losses. The AGV that has highest $P_{us}^{i,k}$ is selected as priority AGV and $P_{e}^{th}$ is chosen according to ~\eqref{eq:ch6:500}. The MCS scheme is selected such that the the error probability for transmitting the current packet is below $P_e^{th}$. The number of RBs required to transmit the packet with selected MCS scheme is allocated to the priority AGV. The error probability is updated with current MCS scheme and $\delta_{i,k}$ is evaluated. If no RBs are available for allocation i.e. $\sum_{i = 1}^N N_{rb,i} = N_{RB}$, the $P_e^{i,k}$ = 1. The control commands are then applied to AGV's actuator depending on $\delta_{i,k}$. If $\delta_{i,k} = 0$, the current control commands are applied, else if $\delta_{i,k} > 0$ the outdated control commands that were successfully received are applied. The AGV attains new position and the error $\epsilon_{i,k}$ is evaluated, that is used to generate new control commands. 
  

\begin{figure}[!t]
\centering
\includegraphics[width=0.5\textwidth]{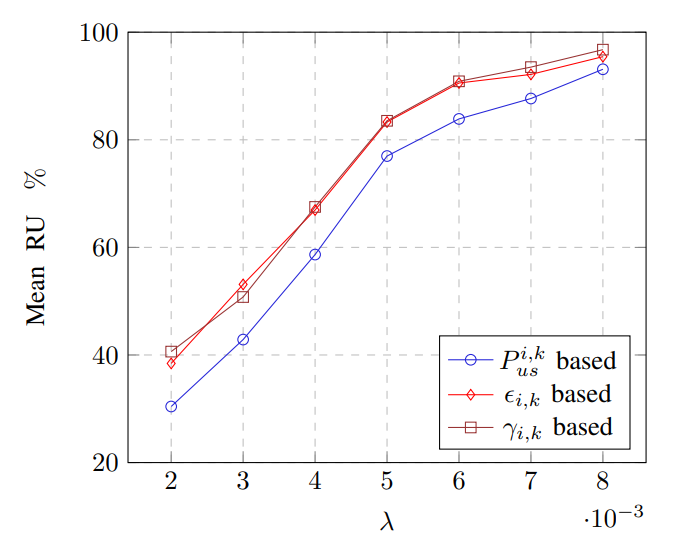}
\caption{Mean Resource utilization over varying AGV arrival rate $\lambda$} 
\label{ch6:result:1} 
\end{figure}




\section{Results} \label{sec:ch6:20}
\subsection{Simulation setup and performance metrics}
AGVs arrive with a Poisson process with arrival rate of  $\lambda = \lbrace 2e^{-3},3e^{-3}, \cdots, 8e^{-3} \rbrace$ and the service time for each AGV is model with negative exponential with rate parameter $\mu = 4e^{-4}$. The Bandwidth  of 1.4MHz is used for downlink transmission with transmit power of 20dBm. The path-loss exponent $\beta$ is 3, the AGVs are distributed within a distance of radius $r$ = 1000 m. The information transmitted at every time-step from an active AGV is 600 bits. Every time-step is of 5ms, consisting of 5-subframes, and 60 RBs per time-step for allocation (6RB/time-slot). The simulation is observed for approx. 4 minutes i.e $5e4$ time-steps for $5$ independent iterations. 

\paragraph{Resource Utilization} The total Resource Utilization (RU) at every time-step is evaluated as 
\begin{align}
RU = \frac{\sum_i^N N_{rb,i}}{N_{RB}} \times 100 \%
\end{align}
where $N_{rb,i}$ is the number of RBs allocated to $i^{th}$ AGV at time-step $k$ and $N_{RB}$ is the total number of RBs available for scheduling. 

\paragraph{Unstable AGVs}
An AGV is unstable if the error at any time-instant is above the error threshold  $\epsilon_{i,k} \geq \epsilon_{th}$, where $\epsilon_{th} = 0.02m$ or the number of consecutive packet loss $\delta_{i,k} \geq n_{max}$, where $n_{max} = 10$ \cite{AGVSCC}. 

\paragraph{Successful AGVs} An AGV is successful if it is stable and completes the track such that at any time-instant $\epsilon_{i,k} < \epsilon_{th}$ and $\delta_{i,k} <  n_{max}$.



\subsection{Resource utilization} The Fig.~\ref{ch6:result:1}, shows the mean resource utilization at every time instant for varying arrival rates of AGVs. The resource utilization and the performance of the AGV is observed under a non-overloaded and an overloaded scenario. In a non-overloaded scenario, sufficient RBs are available for scheduling all the active AGVs. The arrival rate $\lambda$ is low and hence less number of AGVs are active at a given time instant, such that all the active AGVs can be served with sufficient resources. On the contrary in an overloaded case, limited resources are available with more number of active AGVs than the system can ideally support.

\begin{figure}[!t]  
\centering
\includegraphics[width=0.5\textwidth]{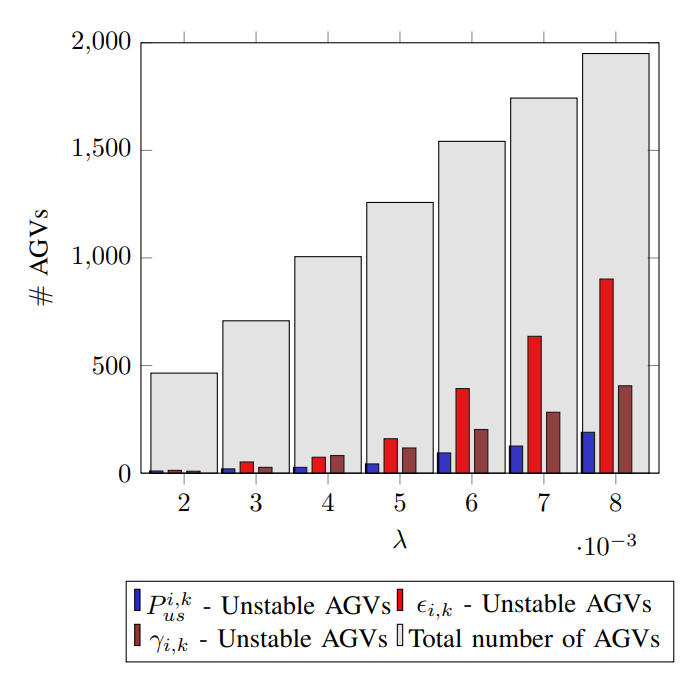}
\caption{Absolute number of Unstable AGVs} 
\label{ch6:result:5}
\end{figure}

\paragraph{Non-overloaded system}  The Fig.~\ref{ch6:result:1} shows the mean RU for different RA schemes that uses $P_{us}^{i,k}$, $\epsilon_{i,k}$ and $\gamma_{i,k}$ as the scheduling metric. The plot below $\lambda < 6 \times 10^{-3}$ shows the resource utilization in an non-overloaded case. The $P_{us}^{i,k}$ based RA scheme requires less resources compared to the SNR based and error based schemes. In $P_{us}^{i,k}$ based scheme the $P_{e}^{th}$ is varied to fulfill the the requirement on probability of instability, hence allocating more RBs to the AGV with higher probability of instability and less RBs i.e. higher MCS to the AGVs that are more likely to be stable. As the number of active AGVs per scheduling time instant increases, $P_{us}^{i,k}$ based outperforms SNR based and error based schemes by utilizing less resources. The $\gamma_{i,k}$ based and $\epsilon_{i,k}$ based schemes have higher resource consumption  as the resources are assigned with constant $P_e^{th}$ and excludes the stability based dynamic resource assignments.  
\paragraph{Overloaded system}
In an overloaded case, for $\lambda > 6 \times 10^{-3}$, the $P_{us}^{i,k}$ based scheme still utilizes less resources compared to $\gamma_{i,k}$ and $\epsilon_{i,k}$ based schemes, but steadily converges at $\lambda = 8 \times 10^{-3}$. The resource utilization for all the schemes approaches \unit[100]{\%}, due to more number of AGVs being activated and hence all the available resources are utilized. However, due to dynamically allocating resources for $P_{us}^{i,k}$ more AGVs can be successfully served even in an overloaded case as shown in Fig.~\ref{ch6:result:5} as compared to $\epsilon_{i,k}$ and $\gamma_{i,k}$ based schemes. Moreover, the $\gamma_{i,k}$ RA scheme shows better performance compared to  $\epsilon_{i,k}$ RA scheme, as less number of AGVs become unstable with similar resource utilization. As the $\gamma_{i,k}$ based RA scheme allocates the resources considering channel conditions, lower is the probability of packet failure. On the contrary control error based RA scheme do not consider channel condition, and allocates more resources to the AGV with higher $\epsilon_{i,k}$ even in the bad channel condition, that leads to packet loss and resource wastage. Therefore, the $\epsilon_{i,k}$ has higher resource utilization and more number of unstable AGVs. 

\subsection{Unstable AGVs}
The Fig.~\ref{ch6:result:5} and Fig.~\ref{ch6:result:6} show the absolute number and percentage of unstable AGVs respectively. In non-overloaded case, $\lambda < 6\times 10^{-3}$ sufficient resources are available hence very few AGVs exceeds $\epsilon_{th}$ and $n_{max}$. Error based resource allocation scheme has the maximum number of unstable AGVs even though the priority is given to the AGV with highest control error. This is due to the allocation of more resources to an AGV without considering its channel condition. Allocating more RBs to an AGV with highest $\epsilon_{i,k}$ might have a higher probability of packet loss, if the AGV is in deep fade. Hence, transmitting the data with lower MCS will not ensure reliable transfer of control information as the received SNR is low to successfully decode a packet. Furthermore, this leads to resource shortage for the AGVs that have good channel condition and comparatively less control error causing higher consecutive packet drops and instability. 

$P_{us}^{i,k}$ based scheme have lower number of unstable AGVs than the $\gamma_{i,k}$ and $\epsilon_{i,k}$ RA scheme as it considers both; the AGV's stability criterion and the channel conditions. If the system prevents consecutive packet loss, less will be the AoI and more accurate control information is applied to the actuators causing less $\epsilon_{i,k}$. Furthermore the $P_{us}^{i,k}$ RA scheme is independent of the control error $\epsilon_{i,k}$ that is evaluated from the position feedback at every time instant via uplink channel. Hence, the $P_{us}^{i,k}$ based RA scheme is robust against the uplink channel conditions and instantaneous control state, but only depends upon the pre-known control parameters like sampling time and velocity of an AGV. 

Also, in an overloaded case, the $P_{us}^{i,k}$ based RA scheme have few number of unstable AGVs as a result of dynamic resource allocation by varying the $P_e^{th}$ such that the $P_{us}^{i,k} < 10^{-9}$. The rate at which the $\%$ of unstable AGVs increase with increase in $\lambda$ is less for $P_{us}^{i,k}$ RA scheme due to higher multiplexing gains. The channel condition based RA schemes like $P_{us}^{i,k}$ and $\gamma_{i,k}$ shows higher gains in an overloaded case when compared to $\epsilon_{i,k}$ RA scheme.

\section{Conclusions}\label{sec:ch6:25}
We proposed a stability based resource allocation scheme for a pragmatic non-linear time-varying closed-loop feedback AGV control system. We evaluated a new scheduling metric 'Probability of Instability' that considers channel condition, consecutive packet losses and control stability for optimally selecting the MCS scheme and allocating the resources. Unlike traditional Maximizing SNR RA scheme, the presented RA scheme allocates more resources to the AGV that is more likely to become unstable. The RA scheme ensures that the probability of instability for all the stable AGVs is below $10^{-9}$ and the control error is maintained within the threshold of \unit[2]{cm}. The performance of RA scheme is analysed and compared with existing SNR based RA scheme and control error based RA scheme. The RA with stability based scheduling metrics utilizes less resources while serving more number of stable AGVs is a system. The AGV's performance and resource consumption is presented for an overloaded and non-overloaded scenario.

\bibliographystyle{IEEEtran}
\bibliography{RA_stability}

\end{document}